**SWSC**

Flares, coronal mass ejections and solar energetic particles and their space weather impacts

 

# The detection of ultra-relativistic electrons in low Earth orbit

Athanassios C. Katsiyannis[1,*], Marie Dominique[1], Viviane Pierrard[2,3], Graciela Lopez Rosson[2,3], Johan De Keyser[2], David Berghmans[1], Michel Kruglanski[2], Ingolf E. Dammasch[1] and Erwin De Donder[2]

[1] Royal Observatory of Belgium, Solar-Terrestrial Centre of Excellence, Avenue Circulaire 3, 1180 Uccle, Belgium
[2] Royal Belgian Institute for Space Aeronomy, Solar-Terrestrial Centre of Excellence, Ringlaan 3, 1180 Uccle, Belgium
[3] Université Catholique de Louvain (UCL), Centre for Space Radiations (CSR), Earth and Life Institute (ELI),
Place Louis Pasteur 3, bte L4.03.08, B-1348, Louvain-La-Neuve, Belgium

**Abstract** − *Aims.* To better understand the radiation environment in low Earth orbit (LEO), the analysis of in-situ observations of a variety of particles, at different atmospheric heights, and in a wide range of energies, is needed. *Methods.* We present an analysis of energetic particles, indirectly detected by the large yield radiometer (LYRA) instrument on board ESA's project for on-board autonomy 2 (PROBA2) satellite as background signal. Combining energetic particle telescope (EPT) observations with LYRA data for an overlapping period of time, we identified these particles as electrons with an energy range of 2 to 8 MeV. *Results.* The observed events are strongly correlated to geo-magnetic activity and appear even during modest disturbances. They are also well confined geographically within the $L = 4$–6 McIlwain zone, which makes it possible to identify their source. *Conclusions.* Although highly energetic particles are commonly perturbing data acquisition of space instruments, we show in this work that ultra-relativistic electrons with energies in the range of 2–8 MeV are detected only at high latitudes, while not present in the South Atlantic Anomaly region.

## 1 Introduction

Relativistic electrons are known to exist at a variety of altitudes since the early days of the space era. During geo-magnetic disturbances (either storms or substorms) the flux of ultra energetic electrons ($J_e$) increases significantly. Many processes have been suggested that can potentially explain the phenomenon, but the lack of observations over many ionospheric areas makes difficult to confine the problem sufficiently. It is well established that high solar wind velocity ($V_{sw}$) in combination with high magnetic activity accelerates electrons to ultra-relativistic velocities (see Paulikas et al. (1979) for the initial work and Reeves et al. (2011) and references therein for the most recent developments). Other space weather parameters, such as the strength of the interplanetary magnetic field (IMF $|\mathbf{B}|$), and the solar wind density ($n_{sw}$), have been shown to correlate well with $J_e$ (Kellerman and Shprits, 2012). Wing et al. (2016) applied advanced information theory techniques to rank the impor-tance of nine such parameters and found $V_{sw}$ as by far the most important parameter, with IMF $|\mathbf{B}|$ a clear second. Kanekal et al. (2001) demonstrated that the acceleration process is of a global scale and the fluxes are uniform over the same $L$ zone. Several authors (see O'Brien et al. (2003), Clilverd et al. (2010), Wing et al. (in press) for the most recent work) have detected energetic electrons precipitating down to the lower ionosphere and the upper atmosphere.

In this paper, we report on a recurrent detection of the ultra-relativistic electrons at the level of LEO by two different science instruments on-board two satellites of the PROBA fleet. Their detections are complementary to those by Lorentzen et al. (2001), O'Brien et al. (2003), Clilverd et al. (2010), and Wing et al. (2013) since the perturbations reported here were made in different geographical locations. One of the main characteristics of the detected electron population is a clear dawn-dusk asymmetry, similarly to what Lorentzen et al. (2001), Stubbs et al. (2001), O'Brien et al. (2003) and Wing et al. (2013) have reported for a variety of particles.

In the next section of this paper, the large radiometer (LYRA) instrument on-board the project for on board autonomy 2 (PROBA2) satellite will be presented. Section 3 contains the profile signature of the reported signal, while Section 4 presents the correlation between the likelihood of a detection and the value of the geo-magnetic $ap$ index. The next section contains the geographical distribution of the perturbations, and Section 6 compares the detections to those made by the energetic particle telescope (EPT) on-board the project for on board autonomy − vegetation (PROBA-V) satellite. Section 7 presents the dawn-dusk asymmetry and provides a possible explanation of the phenomenon. A further discussion on the finding of this work can be found in the last section.

*Corresponding author: katsiyannis@oma.be

**Table 1.** The different detector technologies and wavelength ranges for all LYRA units and channels. The Metal-Semiconductor-Metal (MSM) and p-i-n (PIN) detectors are diamond-based, while the p-n (PN) detectors use conventional silicon bases.

| Channels (bandpasses) | 1: Lyman-α (120–123 nm) | 2: Herzberg (190–222 nm) | 3: Aluminum (17–80 + 1–5 nm) | 4: Zirconium (6–20 + 1–2 nm) |
| --- | --- | --- | --- | --- |
| Unit 1: | MSM (diamond) | PIN (diamond) | MSM (diamond) | PN (silicon) |
| Unit 2: | MSM (diamond) | PIN (diamond) | MSM (diamond) | MSM (diamond) |
| Unit 3: | PN (silicon) | PIN (diamond) | PN (silicon) | PN (silicon) |

## 2 The LYRA instrument

LYRA (Dominique et al., 2013) is an ultraviolet radiometer of the PROBA2 mission, launched by the european space agency (ESA) on 2nd November 2009. It orbits at an altitude range of 707 to 725 km, on a dawn/dusk polar orbit, with an inclination angle of 98.2° and a orbital period of 99.1 min.

LYRA consists of three redundant units that only differ in the detector technology. As radiometers they do not provide angular information but they have high time resolution that can reach the rate of 100 measurements per second (i.e. 100 Hz). Each unit contains four channels with wide bandwidths that expand from UV to soft X-rays. Table 1 contains the approximate bandwidth and the detector technology for each channel of each LYRA unit. The field of view of all channels is 5° and because of the orbital characteristics an eclipse "season" occurs every year from November until February. During this season the observation of the Sun is obscured by the Earth's atmosphere and landmass for a small part of the orbit. Unit 2 is constantly in use throughout the mission while the other two units are only used for calibration purposes and special campaigns.

## 3 The LYRA detections

The continuous LYRA measurements are occasionally affected by events of unknown cause that cannot be traced to a direct solar or instrumental origin. They were as likely to appear during eclipses as during solar observations. Figure 1a contains a characteristic profile of the most common type of such events made by the channel 4 (Zirconium) of unit 2. It consists of a bell-shaped rapid increase and decrease in counts simultaneously with high levels of fluctuation. The whole phenomenon lasts consistently about 100 s and was always observed in the Aluminum and Zirconium channels (i.e. channels 3 and 4). Those perturbations never appeared on measurements taken with the LYRA shutters closed (such measurements are occasionally taken for calibration purposes), which rules out instrumental effects and highly energetic cosmic rays.

The lack of detections from channels 1 and 2 allowed for two distinct possible origins of the signal. Since channels 1 and 2 are sensitive to UV while channels 3 and 4 to EUV and soft X-ray wavelengths (see Tab. 1), one obvious explanation was that photons within the energy range of the Aluminum and Zirconium channels but outside the range of the Lyman-α and Herzberg channels were causing the disturbances. Still, there was another possibility: The optical filters of channels 1 and 2 are made of thick glass while the filters of channels 3 and 4 are made of thin metal. As a consequence, highly energetic particles can have enough momentum to penetrate the filters of the Aluminum and Zirconium channels but not enough for the Lyman-α and Herzberg filters. Distinguishing between those two possibilities is fundamental to understand the nature of the detections.

Given the large number of those events in the LYRA data set, an automatic detection algorithm was therefore implemented. It was based on the most important and prominent characteristics of the above mentioned profile, namely the constant duration, the bell-shape profile and the increased fluctuations. The algorithm consists of the following steps and was applied to all data produced by all four channels of the nominal unit (i.e. unit 2) for the period of 1st January 2010 to 31st December 2015.

- Any perturbation of the time-series caused by known effects was filtered out. Those effects are: spacecraft rotation, occultation events by the Earth, and known solar flares. This breaks the time series into several smaller ones that typically last for a few hours.
- All resulting time-series are smoothed with a window of 100 s since this is the characteristic time scale of the event (see Fig. 1b).
- The smoothed curve is extracted from the original so that only the fluctuation over the Gaussian-like profile remains (see Fig. 1c).
- The mean value and the standard deviation of the time series resulting from step 3 was calculated.
- The number of data points with values of $4\sigma$ above the mean is calculated (represented by a horizontal line in Fig. 1c).
- If the number produced by the previous step exceeds 100 then a detection is credited.
- The duration of the detection is defined as the length of time between the first and the last instance where the signal jumps above the threshold of step five. This duration must be longer than 50 s and shorter than 200 s for a detection to be valid.

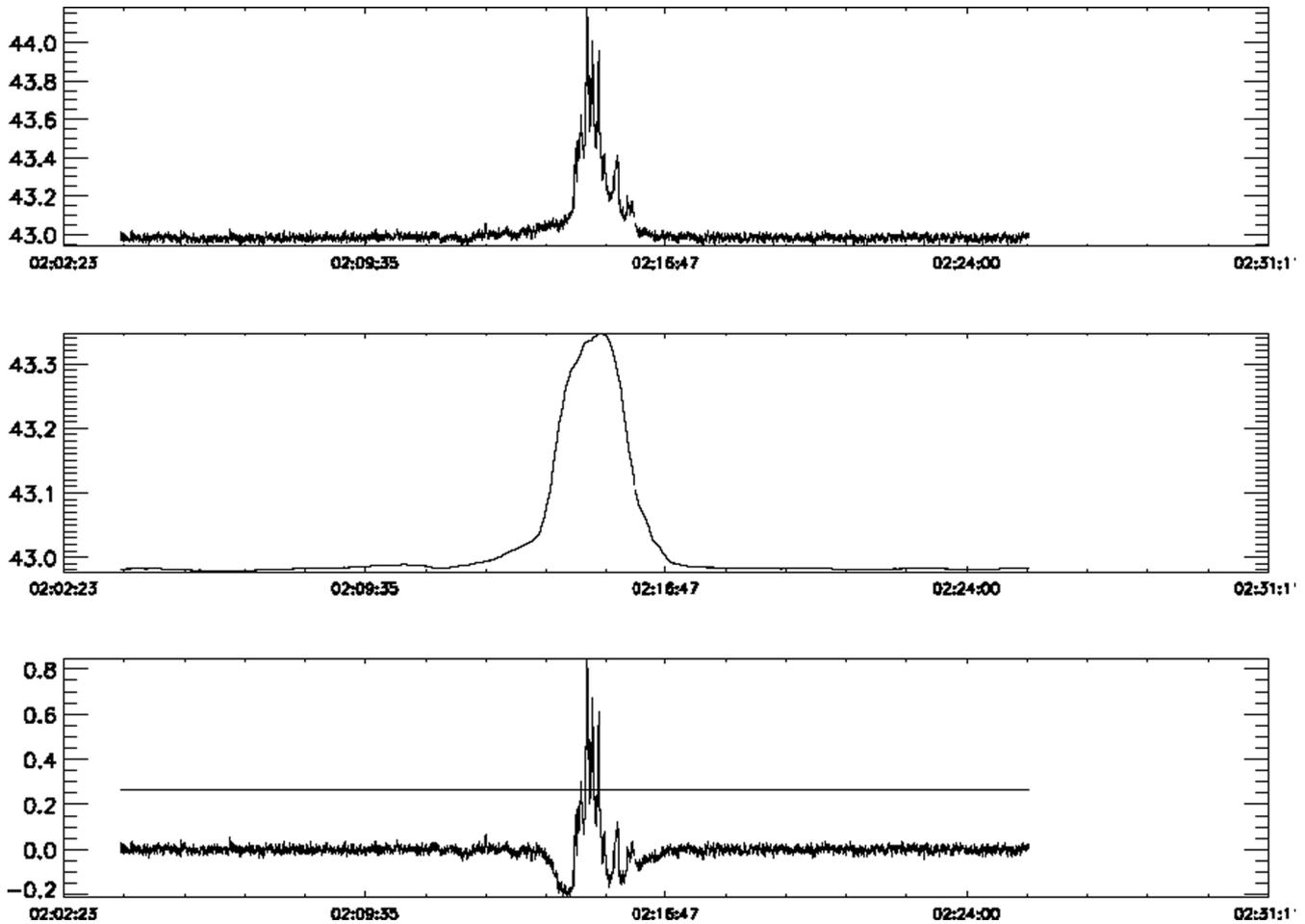

**Fig. 1.** A typical detection made on 9th April 2010 by the channel 4 of unit 2. The time series is filtered with a high-pass filter: Panel A contains the raw data, B the same time series smoothed aver a 100 s window and C the difference between the two. The horizontal line in the third graph corresponds to the $4\sigma$ level of the Gaussian noise.

Approximately 750 detections were identified by the above procedure on each of the Aluminum and the Zirconium channels while none were made for the Lyman-$\alpha$ and Herzberg channels. For more information on the numbering and the approximate wavelength range of the LYRA channels, see Table 1. The detections made by the two channels were systematically almost simultaneous.

## 4 Statistical correlation with the geo-magnetic *ap* index

Close examination of the time of appearance of a sample of detections reveals no obvious connection of the phenomenon with specific solar events. Additionally, manual examination of the data taken during the eclipses of the Sun by the Earth (that occur during the November to February months due to the orbital characteristics of the satellite), shows that the effects also appear during the eclipses. As this proves that the observations are not solar photons, the investigation turned to the ionosphere as the source of the detections.

The *ap* index is a quantified measurement of the disturbance of the horizontal component of the earth's magnetic field. It is a linear scale of integers from 0 (for no disturbance) to 400 (the assumed possible maximum) and it is calculated as the average of a three hour period (i.e. eight averages every day).

The following procedure was used for the investigation of the possible correlation between geo-magnetic activity and likelihood of detections. It is applied to the whole data set analyzed by the automatic algorithm Section 3 (namely, from 1st January 2010 to 31st December 2015):

For any detection made by the automatic algorithm the *ap* value at the time of the event was recorded.

- for any given *ap* value of the index, the number of detections associated with it was summed up;
- a likelihood of detection was computed by normalizing the number of detections happening at one given *ap* index by the total number of occurrences of this *ap* index. For example if a certain *ap* value appears 1000 times between 1st April 2010 and 30th November 2014 and the total number of detections during those 1000 3-hour periods is 2000 then the normalised value will be 2. This value will be called likelihood of detection.

Figure 2 shows the correlation between the *ap* index value and the likelihood of a detection for both channels where detections were found. The correlation is approximately linear and is found to be much stronger than with other indices of ionospheric activity (namely, $D_{st}$ and $AE$). The relatively weak correlation with the $D_{st}$ index is typical of the findings of (Reeves, 1998) on relativistic electron fluxes. Figure 2, on the other hand, shows that even very modest substorms of, say, $ap = 20$ have 10% chance of producing an event. Very high $ap$ values typically correspond to extended periods of extreme space weather, during which the LYRA measurements are often dominated by the direct solar flare signal. This explains the very low likelihood for $ap = 170$ since all periods corresponding to flare events have been filtered out from the data. It is therefore impossible to argue whether those events exist for $ap > 170$.

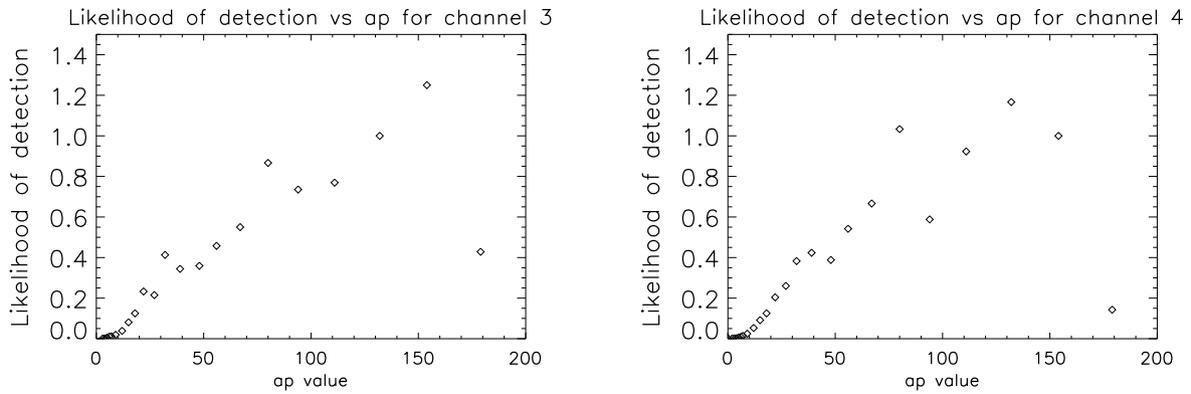

**Fig. 2.** Likelihood of a detection taking place at any three-hour window of a certain *ap* value for channels 3 and 4 of unit 2.

**Table 2.** The characteristics of the L-shell statistics. The spread of values is relatively small, indicating a common geographical origin along the $L^{\sim}5$ zone. The larger spread of values to higher (in comparison to lower) L is due to the geographical proximity of the high-L zones. LYRA channels 3 and 4 detect almost the same events.

| Channels/Statistics | Mean | Median | Min value | Max value |
|---|---|---|---|---|
| Aluminum | 5.8 ± 1.3 (65° ± 3°) | 5.5 (65°) | 3.1 (55°) | 14.6 (88°) |
| Zirconium | 5.6 ± 1.4 (64° ± 3°) | 5.3 (64°) | 1.5 (36°) | 15.5 (88°) |

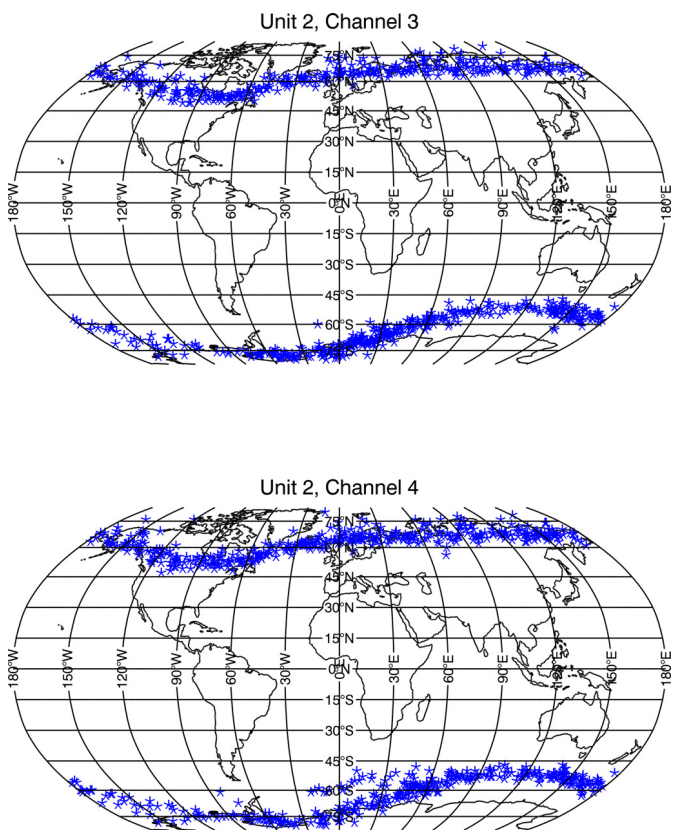

**Fig. 3.** Maps of the LYRA detections for channels 3 (top) and 4 (bottom).

## 5 Geographical distribution of the detected fluctuations

Figure 3 contains maps of the locations of the satellite when detections were made for channels 3 and 4. Table 2 contains the average L-shell ($L$) and invariant latitude ($\Lambda$) values together with their standard deviations for both channels. The difference between the median and average values is due to the non-symmetrical distribution of $L$-values (i.e. the higher the $L$, the closer together the shells are geographically). The consistency between the values of Table 2 is strong evidence that the mechanism that accelerates the electrons is located in a specific physical location along the $L \sim 5$ shell. The 1.3 spread of $L$-values is smaller than the length of travel of the satellite during the 100 s of the typical perturbation duration. Additionally, and as mentioned before, the magnetic field distortion is known to affect the trajectories of electrons very significantly and bring them down to much lower $L$ zones. Similarly, Onsager et al. (2002) found >2 MeV electrons on $L$ zones as low as $L = 4$, concluding that they originated from the geostationary orbit that reached lower $L$ zones by magnetic field distortion.

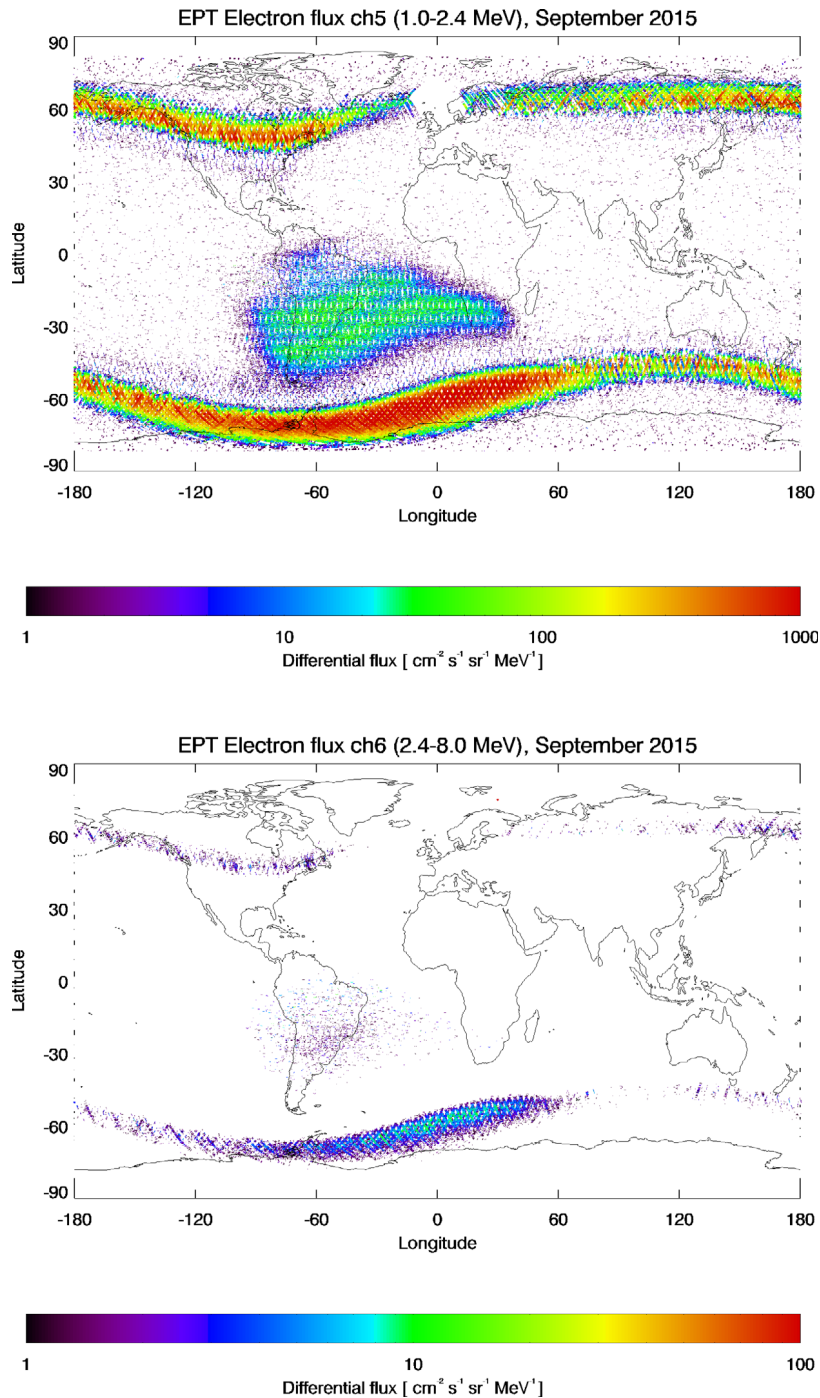

**Fig. 4.** Maps of the EPT electron detections for channels 5 (1–2.4 MeV, top graph) and 6 (2.4–8 MeV, bottom graph). The channel 5 detects electrons at the same locations as LYRA but it also detects electrons in the South Atlantic Anomaly (SAA) area. The channel 6 map is more similar to the maps of Figure 3 since the SAA is far less prominent, while a significant amount of detections remains around the $L = 6$ area.

Additional analysis was performed to determine the existence of a seasonal (i.e. summer versus winter) variation of the likelihood of a detection. The data were divided into two seasons (November to April and May to October) and the numbers of detections in each hemisphere for a given season was compared. No dependence on season or hemisphere was found. A similar comparison of the distribution of the detections along the two hemispheres did not revealed any asymmetry on the $L$ values. Therefore, the appearance of the particles in magnetically conjugate regions in both hemispheres confirms that the phenomenon must be related to charged particles on closed magnetic field lines.

The profile of the disturbances can be described as a smooth, almost Gaussian, increase of the background with transient, very intense spikes overimposed (see Fig. 1). The most obvious explanation of those spikes is either that they are short-lived bursts of electrons, or that the satellite flies over very narrow regions of high activity (or both). As the nominal cadence of LYRA is very high (20 Hz) and occasional observations reach the extremely high value of 100 Hz, it was possible to analyse the relationship between certain statistical characteristics of the signal of the events with the sampling ratio. The characteristics examined were the background and peak values and they were found to be linearly related to the sampling ratio (i.e. doubling the exposure time doubled both values). This result indicates that the observed phenomenon is either of a continuous nature or the bursts happen in either a much shorter time scale (i.e. much less than 100 Hz), or shorter spatial scale (i.e. much less than 75 m that corresponds to 10 ms of flight).

## 6 The EPT detections

The EPT is a new instrument that provides high-resolution measurements of the charged particle radiation environment in space performing with direct electron, proton and heavy ion discrimination (Cyamukungu et al., 2014a, b). It was launched on 7 May 2013 to a polar LEO at an altitude of 820 km on-board the ESA satellite PROBA-V. The orbital characteristics are very similar to those of PROBA2, namely, a period of 101 min and an inclination of 98.73° and 10:30 am as nominal local time (LT) at the descending node (Pierrard et al., 2014). The detector measures the particle fluxes for seven virtual channels for electrons from 500 keV to 20 MeV, 11 channels for protons from 9.5 to 248 MeV and 11 channels for Helium ions from 16 MeV to 1 GeV. To gain information about the nature of LYRA perturbations, we compared them to the observations of EPT, as a more specialised instrument to in-situ detections of ultra-relativistic electrons.

An example of electron map obtained with the EPT instrument is illustrated on Figure 4 corresponding to the data observed during the month of September 2015. The upper panel shows channel 5 from 1 to 2.4 MeV and the bottom panel illustrates channel 6 from 2.4 to 8 MeV. High electron fluxes are observed at high latitudes and correspond to the penetration of the outer radiation belt at low altitude. These fluxes are enhanced during geo-magnetic storms as observed several times during the highly active 2015 year (Pierrard and Lopez Rosson, 2016). No protons are observed at these high latitudes. Energetic electrons (and protons) are observed in the South Atlantic Anomaly (SAA) by Lopez Rosson and Pierrard (2017), but for energies >2.4 MeV the fluxes are only observed at high latitudes and not in the SAA (as it is also the case in LYRA observations). Van Allen Probes observations have shown that electrons above 2 MeV are not present in the inner belt and in the SAA, but they are well present in the outer belt (Baker et al., 2014). This indicates that LYRA detected energetic electrons with E>2 MeV. The gap in electron detections above the North sea (at an approximate longitude of 0°) is due to the interruption of EPT data acquisition during the connection of the PROBA-V satelite with the ground station for data transmission (see Cyamukungu et al. (2014a) for more details).

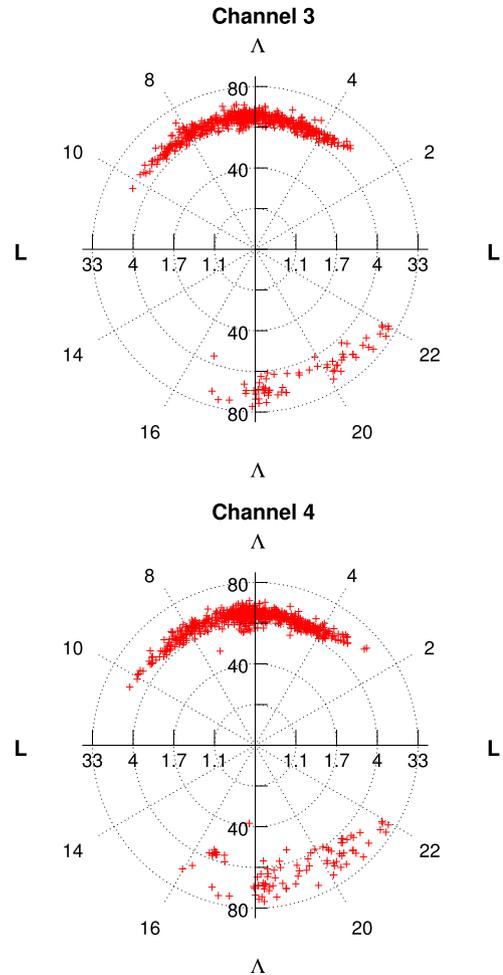

**Fig. 5.** The locations of the detections in magnetic local time (*MLT*) versus both invariant latitude (*Λ*) and L-shell (*L*) coordinates for LYRA channels 3 (top) and 4 (bottom). The radius r is proportional to *Λ* and the angle *φ* to the *MLT*. The y-axis is marked with invariant latitude, while the x-axis is in *L*-shell values. The red crosses correspond to the location of the detections, while the dotted circles correspond to different *Λ* values and the dotted lines note the various *MLT*s.

## 7 Dawn-dusk and day-night asymmetries

To investigate further the origin of the perturbations, the possibility of asymmetries in the distribution was investigated based on two criteria: The north versus south hemispheres distribution and the effect of the *ap* value to the geographical distribution. On both cases no statistically significant difference was found. However, a very large dependance was found on the LT of the satellite at the time of the detection. More specifically, 646 (686) detections were made during dawn, while 74 (97) detections where made in dusk for channel 3 (the numbers in parenthesis are the detections for channel 4).

Since this phenomenon is obviously magnetic, and in order to investigate further the dawn-dusk asymmetry found in LT, the magnetic local time (*MLT*), the L-shell (*L*), and the invariant latitude (*Λ*) of all the detections were calculated using the space environment information system (SPENVIS), described by Heynderickx et al. (2000). The location of the detections in *MLT* versus *L* and *Λ* coordinates is plotted in Figure 5, where a strong dependence of the detections on *MLT* can be clearly seen. To correct for bias in the *MLT* and $\Lambda/L$ coordinates caused by the satellite's orbit, the number of passes of the PROBA2 satellite for each individual *MLT* (in beams of 1 h) was calculated. Those passes are defined as the number of 30 s periods that the satellite spend on each *MLT* and with $\Lambda \in (30°, 80°)$ (approximately $1.3 < L < 33$). Additionally, the number of detections per *MLT* (also in beams of 1 h) was calculated and the results are plotted as a histogram for each channel in Figure 6. To correct for the effects of the orbit, the number of detections per *MLT* was divided by the number of satellite passes through the same *MLT* and the results are also displayed in Figure 6. As expected from a satellite on a dawn-dusk orbit, there is a significant bias on the distribution of detections, but it doesn't not alter the overall phenomenon qualitatively.

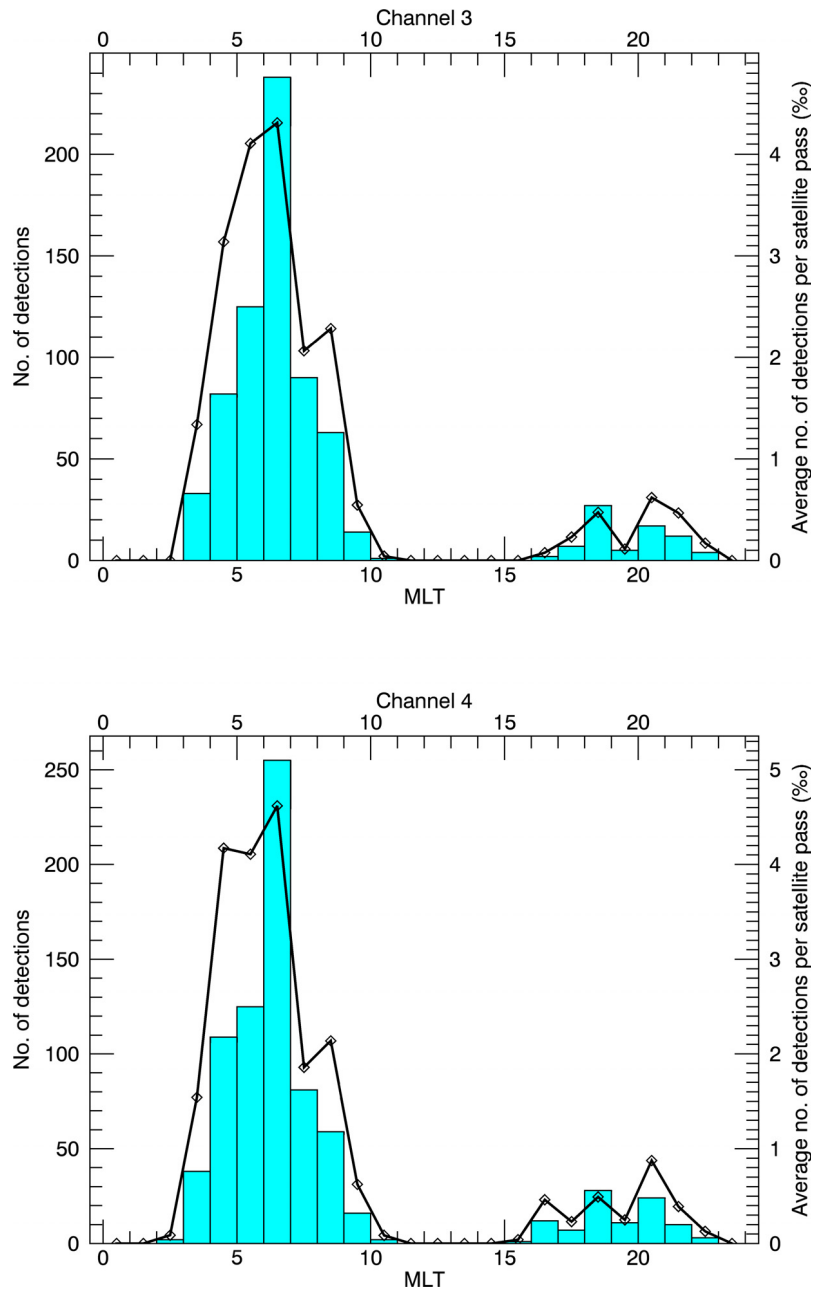

**Fig. 6.** The distribution of LYRA detections and the likelihood of a detection for each satellite pass, along all *MLT*s. The number of detections is presented as a histogram with the scale on the left y axis. The line corresponds to the likelihood of a detection, i.e. the number of detections divided by the number of passes of the PROBA2 satellite over the same $\Lambda$ range ($30° < \Lambda < 80°$). The right y axis contains the scale of the likelihood of a detection per satellite pass and the units are detections per thousand passes (‰).

Another distinct difference of the dawn and dusk populations, which is clearly seen in Figure 5, is their statistical distribution along the $\Lambda$ axis. Table 3 contains the basic statistical characteristics of the two distributions, where it is obvious that the dawn detections have a marginal smaller average $\Lambda$ than the dusk, but the spread of $\Lambda$ values is significantly smaller in dawn than dusk.

**Table 3.** Basic statistical characteristics of the distribution of detections along the invariant latitude ($\Lambda$). The values without brackets correspond to channel 3, while those with to channel 4 of LYRA.

| MLT statistics | Mean (°) | Median (°) | Min (°) | Max (°) |
|---|---|---|---|---|
| Dawn (00:00-12:00): | 65 ± 2 (64 ± 2) | 65 (64) | 58 (49) | 74 (71) |
| Dusk (12:00-24:00): | 69 ± 4 (66 ± 7) | 69 (67) | 55 (36) | 88 (88) |

**Table 4.** The number of detections versus the *MLT* for both LYRA channels and the likelihood of a detection on each channel, after correcting for bias by the satellite orbit. The units of the likelihood are the average number of detections per 1000 passes (‰). The numbers outside parenthesis are for channel 3 and those inside for channel 4.

| MLT range | Early dawn [00:00–06:00] | Late dawn [06:00–12:00] | Early dusk [12:00–18:00] | Late dusk [18:00–24:00] | Average per hour [00:00–24:00] |
|---|---|---|---|---|---|
| No. of detections | 240 (274) | 406 (413) | 9 (20) | 65 (76) | 30 ± 56 (33 ± 59) |
| Likelihood (‰) | 1.62 (1.86) | 2.05 (2.09) | 0.06 (0.13) | 0.32 (0.38) | 0.8 ± 1.3 (0.9 ± 1.4) |

Additionally, in Figure 5, there is a lack of detections at 11:00–16:00 and 23:00–02:00 *MLT*. This effect further splits the detections into four distinct groups: Early dawn, late dawn, early dusk, and late dusk. Table 4 contains the number of perturbations for those four cases. Any acceptable explanation of the phenomenon should explain both characteristics: The dependence of the number of events with *MLT* and the constraint to the displayed range of $\Lambda$ values. It should be noted here that there is no statistically significant difference between the spread of $\Lambda$ values between day and night, unlike the dawn-dusk difference in spread mentioned previously. It is also apparent from Figure 6 that the satellite's orbit has a significant effect on both the dawn-dusk and the night-day asymmetries, but this does not alter the results qualitatively in regard to the dawn-dusk differences. Furthermore, it should be noticed that there is a statistically significant number of PROBA2 passes by all the *MLT* range, but the orbit is restricted to higher $\Lambda$s for those *MLT* areas with no detections. As such, the lack of detections in the above mentioned areas is probably due to PROBA2's orbit.

Figure 7 displays the maximum fluxes detected by EPT for the period of May 2013 to December 2015 in the same coordinate system as Figure 5. Although it is obvious that there is no significant overlap of orbits in the *MLT* vs $\Lambda$ space for the two satellites, certain comparisons can be made:

– the intensities observed by the EPT are much more balanced between the two hemispheres than the detections by LYRA;
– although the spread of the LYRA detections in the $\Lambda$ dimension is much smaller in the dawn sector than the dusk, in the EPT data the fluxes remain equally concentrated in $\Lambda$ regardless of the *MLT*;
– the most intense flux detected by EPT in the *MLT* is at lower $\Lambda$ than the LYRA detections. This is equally valid for the 08:00–10:00 and 20:00–22:00 *MLT* regions where the orbits of the two satellites overlap.

One feasible explanation of the above mentioned differences is that the LYRA detections are caused by electrons with an energy restricted to a subrange of EPT's channel 6 energy window. This assumption is in good agreement with the lack of electron detections by LYRA when flying over the SAA (as also noted in Sect. 6), since the higher energy electrons are known not to penetrate the SAA region. It is, therefore, possible that LYRA's energy range is much higher than 2.4 MeV, but still below the 8 MeV limit of EPT's channel 6. The latter is evident as there is no correlation between the LYRA detections and the detections of electrons by EPT's channel 7 (which is sensitive energies above 8 MeV). A factor of twenty in the cadence of the two instruments (0.5 Hz for EPT and 10 Hz for LYRA) is a second important difference. As such, the high amplitude and short duration peaks that are clearly resolved in the LYRA data, are smoothed out in the EPT signal. Overall, one might argue that EPT observes the same electrons as LYRA, but that their fluxes only marginally contribute to the EPT signal as they constitute only a small part of the observe population. A more detailed comparison of the detections of the two instruments may shed more light on this hypothesis, but the small size of the common *MLT*-$\Lambda$ areas, and the different timing of the passes of the two satellites make such comparison difficult.

Dawn-dusk asymmetries are not new to ionospheric physics. Measurements of relativistic electrons by Onsager et al. (2002) revealed a strong correlation between LT and electron fluxes and the depletion of >2 MeV electrons from the magnetosphere. They found a strong correlation between LT and >2 MeV electron flux and a significant difference on the timing of the depletion as a function of LT (with the dropouts occurring first at dawn and then at dusk). Their detections were made over a very similar range of *L*-shell values (with measurements being in-situ to the apex of the *L*-shells), but their emphasis was on the timing of the flux variations and they did not provide the dawn/dusk detection ratio. Instead, they studied the relation between the LT and timing of the electron dropouts, concluding that they are caused by localised changes that are difficult to determine with certainty.

The profile of the perturbations of the time series (see Fig. 1) is compatible with the microbursts first reported by Anderson and Milton (1964) and many other authors (for

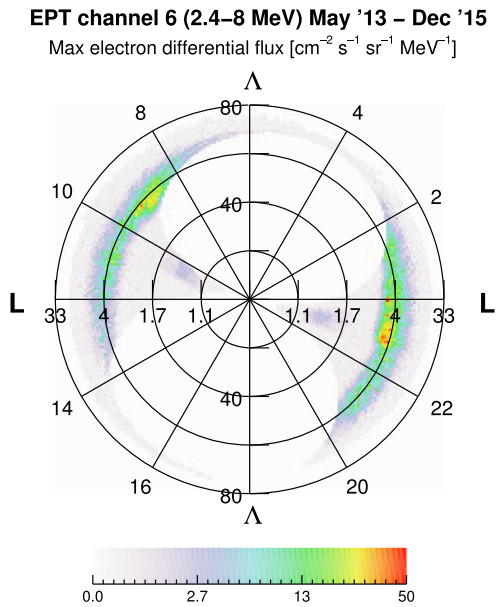

**EPT channel 6 (2.4–8 MeV) May '13 – Dec '15**

Max electron differential flux [cm⁻² s⁻¹ sr⁻¹ MeV⁻¹]

**Fig. 7.** The maximum electron intensities recorded by channel 6 of the EPT/PROBA-V instrument plotted on a *MLT* vs *A/L* polar diagram. Similarly to the polar diagram of Figure 5, the radius r is proportional to *Λ* and the angle *φ* to the *MLT*. The y-axis is marked with invariant latitude, while the x-axis is in *L*-shell values. The flux scale is logarithmic, where the white areas are outside PROBA-V's orbit, the faint purple is noise.

e.g. Nakamura et al. (2000); Lorentzen et al. (2001); O'Brien et al. (2003) and references there in. Nevertheless, important differences exist between the geographical distribution of the detections reported by this paper and previous works. Lorentzen et al. (2001) reported the peak of the flux of electrons with energies >2 MeV at the region of $4 < L < 6$ (60°–65° invariant latitude) and $03:00 < MLT < 09:00$, while O'Brien et al. (2003) at the same L region as Lorentzen et al. (2001) but different MTL ($06:00 < MLT < 12:00$). In comparison, and as seen in Figure 5, the highest number of LYRA detections are around the 60°–70° invariant latitude ($4 < L < 8$) and $03:00 < MLT < 09:00$. It is, therefore, apparent that the Lorentzen et al. (2001) results are closer to the detections reported here, at least for the *MLT* dawn section. As far the *MLT* dusk section is concerned, the LYRA results show a high probability in the range of 60°–80° invariant latitude ($4 < L < 30$) and $18:00 < MLT < 22:00$, while Lorentzen et al. (2001) reported a lower invariant latitude (60°–65°) and a significantly different *MLT* range ($19:00 < MLT < 24:00$).

Whistler mode chorus waves have been found (Meredith et al., 2001) to accelerate electrons to relativistic speeds in the region of $3 < L < 7$ for a variety of *MLT*. More specifically, low latitude ($|\lambda_m| < 15°$, also called equatorial) chorus waves are found to drift from midnight *MLT* to noon *MLT*, while the higher latitude (i.e. $|\lambda_m| < 15°$) whistler mode waves are found to affect the 06:00–15:00 *MLT* region. As such, the early and late dawn detections reported here are in full agreement with the hypothesis that chorus waves can accelerate electrons to ultra-relativistic speeds (with equatorial waves causing the early dawn and high latitude waves causing the late dawn to noon detections). The "desert" between the two groups at around 06:00 LT is caused by the lack of magnetic field lines from the appropriate *L* zone.

Electron cyclotron harmonic (ECH) waves (Meredith et al., 2000) are also a possible explanation for the detections made in the early dawn section of Figure 5. They appear stronger at the dawn section and for $6.0 \le L < 7.0$ or $3.8 \le L < 6.0$ (depending on the circumstances), which matches well the perturbations reported here. Nevertheless, as the ECH waves affect the same range of LT and *L* as the whistler mode chorus, there is no obvious way to distinguish the electrons accelerated by those two types of waves, and thus, no reliable way to determine if any events are caused by ECH waves.

Another possible explanation for the early dusk detections are the electromagnetic ion cyclotron (EMIC) waves that undergo cyclotron resonant interactions of a multi-ion plasma with relativistic electrons in the outer radiation belt (for an explanation of the mechanism see Summers and Thorne (2003) and references therein). Meredith et al. (2003) observed electrons accelerated by such waves and showed their prevalence in the dusk sector (mostly 14:00–22:00 *MLT*). Similarly to the detections presented here, they found most of the EMIC waves in the early dusk sector (14:00–18:00 *MLT*) and for $3 \le L < 7$. Although they focused on energies below 2 MeV and their wave detections could extent only to $L=7$, EMIC waves are considered capable of accelerating electrons to energies larger than 2 MeV (Summers and Thorne, 2003).

The last group of detections to be associated with an electron acceleration mechanism is the late dusk (i.e. 18:00–24:00 LT). Wing et al. (2013) and Wing et al. (in press) have detected "monoenergetic" electrons predominantly on the 18:00–24:00 *MLT* sector, using the SSJ4 and SSJ5 detectors on-board the DMSP F12 and F15 missions. Although the sensitivity of the SSJ4/5 instruments does not extend to the range of electron energies reported here, there is a strong LT and *L*-zone correlation with the LYRA-EPT detections. Wing et al. (2013) and Wing et al. (in press) suggested two possible mechanisms that explain this acceleration. First, the stretching of the magnetotail during the growth phase of a storm (this causes the intensification of the field currents: see Watanabe and Iijima (1993) and references their in). The second is the kinetic-ballooning/interchange mode in the magnetotail (such as the one discussed by Pritchett and Coroniti (2010)), that is associated with interchange heads that generate aurora streamers. The benefit of the second possibility is the association of the aurora streamers with fast ionospheric flows (Nakamura et al., 2001; Sergeev et al., 2004), which are in turn associated to low frequency Alfvén waves (Damiano and Johnson, 2012). This interaction of kinetic-ballooning/interchange oscillations, fast flow and Alfvén waves best fits the time profile of the Wing et al. (in press) observations.

In any case the loss of electrons from the dusk sector of the magnetic tail is well documented by Green et al. (2004). It is worth mentioning at this point that the electrons reported here are permanent or temporal losses from the magnetosphere (although, the LEO orbit of the PROBA2 satellite makes it likely that the electrons are permanently lost). Also, since the satellite only passes through the affected area every ~100 min (i.e. the duration of one orbit), it is impossible to determine if the events of the four groups (i.e. early dawn, late dawn, early dusk, late dusk) happened simultaneously or not.

## 8 Discussion

Microbursts of relativistic electrons are known to exist and have been reported by many authors (see Nakamura et al. (1995) as an example), with Nakamura et al. (2000) having shown a correlation of microbursts and the dawn side of the plasmasphere. Lorentzen et al. (2001) and O'Brien et al. (2003) reported that MeV microbursts are usually observed by the SAMPEX spacecraft simultaneously with chorus waves detected by the POLAR satellite (for a description of the HILT/SAMPEX and PWI/POLAR instruments, see Klecker et al. (1993) and Gurnett et al. (1995)). We are reporting of a new detection with the LYRA instrument on-board the PROBA2 spacecraft. These observations are complementary to other observations of MeV electrons that were made at orbits with different characteristics, (including those of EPT/PROBA-V, SAMPEX and POLAR) as it can be seen in Pierrard et al. (2014), Lorentzen et al. (2001), Stubbs et al. (2001), Onsager et al. (2002), O'Brien et al. (2003), Green et al. (2004), to name a few. As such, it is likely that known dawn-dusk asymmetries in the chorus waves of the magnetosphere (O'Brien et al., 2003) are responsible for the dawn-dusk asymmetry observed by LYRA (an idea also suggested Lemaire [private communication]). Either way, the "spiky" profile of the detections seen in Figure 1 is observed in LYRA acquisitions of any cadence, up to the maximum of 100 Hz.

Extremely low frequency (ELF), ultra low frequency (ULF), and very low frequency (VLF) waves are known to accelerate electrons to energies of several MeV (O'Brien et al., 2003). More specifically, MeV microbursts are considered to be caused by the interaction of chorus waves with trapped electrons (see Horne and Thorne (2003) and references therein). Thus, O'Brien et al. (2003) used microbursts as proxies for VLF/ELF chorus waves and combined them with ULF observations to argue that ULF waves accelerate electrons at L shells higher than the geostationary orbit, while VLF/ELF waves dominate the acceleration at the L shells reported here. Since there are no LYRA detections at higher L shells, and in agreement with O'Brien et al. (2003), we can conclude that only the VLF/ELF waves are known to accelerate electrons to the energies observed by LYRA.

To further investigate the possible effect of the wave-particle interactions to the LYRA detections, data from a variety of other instruments can be utilised. The Van Allen Probes (aka RBSP) mission has a suite of instruments (such as the electric field and waves suite, the electric and magnetic field instrument suite and integrated science, and the relativistic electron proton telescope) that measure both waves and relativistic electrons at the same time as the LYRA and EPT observations. Due to the large number of LYRA detections (over 700 for each of the two channels) and the continuous observations of the EPT and the RBSP instruments, statistical analysis will be required for a comparative analysis. Although this is most likely an important step to better understand the phenomenon, it goes beyond the scope of this paper. This paper presents the LYRA events, explain them as the detection of ultra-relativistic electrons by comparing them to EPT data, and present the basic statistical properties of those detections. A full explanation and understanding of the acceleration mechanism of those detections is kept for future work.

In any case, the perturbations reported in this paper are likely to be electrons of energy range higher than usually observed, with a time profile, and *MLT* versus *Λ* distribution not often found. Those properties are not easily explained by a single model and are, most likely, due to various different mechanisms of electron acceleration that apply to different *MLT*. Thus, the explanations proposed here are only a first attempt and a full modeling of all the observed properties requires extensive additional work. The detections reported by this paper can also provide additional constrains to existing models, as they provide information on the behavior of electrons in an energy range not frequently observed.

*Acknowledgements.* The authors would like to thank Joseph Lemaire for useful discussions. This work was made possible thanks to the Solar-Terrestrial Centre of Excellence, a collaborative framework funded by the Belgian Science Policy Office. LYRA is a project of the Centre Spatial de Liège, the Physikalisch-Meteorologisches Observatorium Davos and the Royal Observatory of Belgium funded by the Belgian Federal Science Policy Office (BELSPO) and by the Swiss Bundesamt für Bildung und Wissenschaft. The authors would like to thank the Scientific Federal Policy Office for funding the P7/08 CHARM project from the Inter-University Attraction Pole framework program. This work was also funded by European Unions Seventh Programme for Research, Technological Development and Demonstration under Grant Agreement No 284461 − Project eHeroes(www.eheroes.eu). M. Dominique acknowledges the support from the Belgian Federal Science Policy Office through the ESA-PRODEX programme. ESA's SPENVIS software was used for calculating the geomagnetic coordinates for all the observations reported here. The editor thanks Sebastien Bourdarie and an anonymous referee for their assistance in evaluating this paper.

---